\documentclass[amsmath,amssymb]{revtex4}
\usepackage{graphicx}
\usepackage{dcolumn}
\usepackage{bm}
\usepackage{color}
\usepackage{bm}
\usepackage{subeqnarray}
\usepackage[active]{srcltx}
\usepackage{amsmath}

\def\|{|\!|}

\def\beq{\begin{equation}}
\def\eeq{\end{equation}}

\newcommand{\ket}[1]{\left | #1 \right \rangle}
\newcommand{\bra}[1]{\left \langle #1 \right |}

\definecolor{darkred}{rgb}{.8,0,0}

\definecolor{darkblue}{rgb}{0,0,.7}

\begin{document}

\title{Local-time representation of path integrals}
\vspace{-2mm}

\author{Petr  Jizba$^{1,}$}
 \email{p.jizba@fjfi.cvut.cz}
 \author{V\'{a}clav Zatloukal$^{2,}$}
 \email{zatlovac@fjfi.cvut.cz}
\hspace{3mm}
\affiliation{\hspace{1mm}\\ $^{1,2}$FNSPE, Czech Technical University in Prague,
B\v{r}ehov\'{a} 7, 115 19 Praha 1, Czech Republic\\ \\
$^1$ITP, Freie Universit\"{a}t in Berlin, Arnimallee 14, D-14195 Berlin, Germany \\ \\
$^2$Max Planck Institute for the History of Science, Boltzmannstrasse 22, D-14195 Berlin, Germany}

\begin{abstract}
\vspace{3mm}
\begin{center}
{\bf Abstract}\\[2mm]
\end{center}
We derive a local-time path-integral representation for a generic 
one-dimensional time-independent
system.
In particular, we show how to rephrase the matrix elements
of the
Bloch density matrix as a path integral over $x$-dependent
local-time profiles. The latter quantify the time that the sample
paths $x(t)$ in the Feynman path integral spend in the vicinity of
an arbitrary point $x$. Generalization of the local-time
representation that includes arbitrary functionals of the local time
is also provided. We argue that the results obtained represent a
powerful alternative to the traditional Feynman--Kac formula,
particularly in the high and low temperature regimes. To illustrate this point, 
we apply our local-time representation to analyze the asymptotic
behavior of the Bloch density matrix at low temperatures.
Further salient issues, such as 
connections with the Sturm--Liouville theory and the Rayleigh--Ritz variational 
principle are also
discussed.

\end{abstract}

\maketitle

\section{Introduction}

The path integral (PI) has been used in quantum physics since the
revolutionary work of Feynman~\cite{Feynman1948}, although the basic
observation goes back to Dirac~\cite{Dirac33,Dirac35} who appreciated the
r\^{o}le of the Lagrangian in short-time evolution of the wave
function, and even suggested the time-slicing procedure for finite,
i.e., non-infinitesimal, time lags. Since then the PI approach
yielded invaluable insights into the structure of quantum
theory~\cite{FeynmanHibbs} and provided a viable alternative to the
traditional operator-formalism-based
canonical quantization. During
the second half of the 20th century, the PI became a standard tool
in quantum field theory~\cite{Ramond} and statistical
physics~\cite{Zinn}, often providing the easiest route to derivation
of perturbative expansions and serving as an excellent framework for
(both numerical and analytical) non-perturbative
analysis~\cite{Kleinert}.

Feynman PI has its counterpart in pure mathematics, namely, in the
theory of continuous-time stochastic processes~\cite{RevuzYor}.
There the concept of integration over a space of continuous
functions (so-called fluctuating paths or sample paths)
had been introduced by Wiener~\cite{Wiener1923} already in 1920's
in order to represent and quantify the
Brownian motion. Interestingly enough, this  so-called Wiener integral
(or integral with respect to Wiener measure) was formulated 2 years before
the discovery of the Schr\"{o}dinger equation and 25 years before
Feynman's PI formulation. 

%

The \emph{local time} for a Brownian particle (in some literature also
called \emph{sojourn time}) has been of interest to physicists and
mathematicians,
since the seminal work of Paul~L\'{e}vy in
1930's~\cite{Levy1939}. In its essence, the local time characterizes the time 
that a sample trajectory $x(t)$ of a given stochastic process spends in the
vicinity of an arbitrary point $X$. This in turn defines a sample
trajectory $L^X$ of a new stochastic process. A rich theory has
been developed for local-time processes that stem from 
diffusion processes (see, e.g., Ref.~\cite{Marcus} and citations therein). 
For later convenience, we should particularly highlight the Ray--Knight theorem which states that the local 
time of the Wiener process can be expressed in terms of the
squared Bessel process~\cite{Borodin,Ray,Knight}. 
In contrast to mathematics, the concept of the local time is not uniquely settled in physics literature. 
Various authors define
essentially the same quantity under different names (local time, occupation time, traversal time, etc.),
and with different applications in mind.
For example, in Ref.~\cite{Sokolovski} the \emph{traversal time} is used to study quantum scattering
and tunneling processes, in~\cite{Paulin} the small-temperature behavior of the equilibrium
density matrix is analyzed with a help of the \emph{occupation time}, while in
Ref.~\cite{Luttinger1982} the large-time behavior of path integrals that
contain functionals of the \emph{local time} is discussed. 

The aim of this paper is to derive a local-time PI representation of the Bloch density matrix, i.e., the
matrix elements $\bra{x_b} e^{-\beta \hat{H}} \ket{x_a}$ of the Gibbs operator.
This can serve not only as a viable alternative to the commonly used Feynman--Kac representation
but also as a powerful tool for extracting  both large and small-temperature
behavior. Apart from 
the general theoretical outline, 
our primary focus here  will be 
on the low-temperature behavior which is technically more challenging than the large-temperature regime.  
In fact, the large-temperature expansion was already treated in some detail in our previous paper~\cite{JizbaZatl}. 
Last but not least, we also wish to promote the concept of the local time which is not yet
sufficiently well known among the path-integral practitioners.

The structure of the paper is as follows. To set the stage we recall in the next section
some fundamentals from the Feynman PI which will be needed in later sections. In Section~\ref{sec:Heuristic}
we provide motivation for the introduction of a
local time, and construct a heuristic version of the local-time representation of PI's. 
The key technical part of the article
is contained in Section~\ref{sec:LT}, where we derive by means of the replica trick the local-time representation of the 
Bloch density matrix. Relation to the Sturm--Liouville theory is also highlighted in this context.
A local-time analog of the Feynman--Matthews--Salam formula~\cite{LvW,BJV} is presented in Section~\ref{sec:Functionals} 
and its usage is illustrated with a computation of the one-point distribution of the local time. 
Since a natural arena for local-time PI's is in thermally extremal regimes, we confine our attention  
in Section~\ref{sec:Asymp}  on large- and small-$\beta$ asymptotic behavior of the Bloch density matrix. There we also
derive an explicit leading-order behavior in large-$\beta$ (i.e.,
low-temperature) expansion. The analysis is substantially streamlined by using 
the Rayleigh--Ritz variational principle.   
Finally, Section~\ref{sec:Conclusion}
summarizes our results and discusses possible extensions, applications, and future developments of the present work.
For the reader's convenience the paper is supplemented with appendix which clarifies
some finer technical details.

\section{Path-integral representation of the Bloch density matrix }
\label{sec:PI}

Consider a non-relativistic one-dimensional quantum-mechanical system described by a
time-independent Hamiltonian $\hat{H} = \frac{\hat{p}^2}{2M}+V(\hat{x})$ where $\hat{p} \ket{x}
= -i \hbar \partial_x \ket{x}$. Throughout this paper we will study the matrix elements
\begin{equation} \label{rhoDef}
\rho(x_a,x_b,\beta) \ \equiv \ \bra{x_b} e^{-\beta \hat{H}} \ket{x_a}\, ,
\end{equation}
of the Gibbs operator $e^{-\beta \hat{H}}$, where
$\beta = 1/(k_B T)$ is the inverse
temperature and $k_B$ is the Boltzmann constant. The matrix $\rho(x_a,x_b,\beta)$, known also as the Bloch density matrix, 
is a fundamental object in
quantum statistical physics, as the expectation value of an operator $\hat{O}$
at the temperature $T$ can be written in the form
\begin{eqnarray}
\langle \hat{O} \rangle \ = \ \frac{1}{Z} \int_{\mathbb{R}}
\int_{\mathbb{R}} dx_a dx_b \ \! \ \! \rho(x_a,x_b,\beta) \ \! \langle x_b| \hat{O} |x_a \rangle\, ,
\end{eqnarray}
where $Z = \int_{\mathbb{R}} dx \ \!\rho(x,x,\beta)$ is the partition function
of the system. In case of need, ensuing quantum mechanical
transition amplitudes can be obtained from (\ref{rhoDef}) via a Wick rotation which formally amounts to the substitution
$\beta \rightarrow i
t/\hbar$, converting thus the Gibbs operator $e^{-\beta \hat{H}}$ to the quantum evolution operator $e^{-i
t \hat{H}/\hbar}$.

The matrix elements in Eq.(\ref{rhoDef}) can be represented via the
path integral as~\cite{FeynmanHibbs,Kleinert}
\begin{equation} \label{PIx}
\rho(x_a,x_b,\beta) \ = \
\int_{x(0)=x_a}^{x(\beta\hbar)=x_b} \!\!\!\mathcal{D}x(\tau) \exp\left\{-\frac{1}{\hbar} \int_{0}^{\beta\hbar}
\!\!\!d\tau \left[\frac{M}{2}\dot{x}^2 + V(x) \right] \right\}\, .
\end{equation}
This represents a ``sum" over all continuous trajectories $x(\tau)$,
$\tau \in [0,\beta\hbar]$, connecting the initial point $x(0)=x_a$
with the final point $x(\beta\hbar)=x_b$. 
It should be noted that the integral $\int_{0}^{\beta\hbar}
\!\!\!d\tau \left[\frac{M}{2}\dot{x}^2 + V(x) \right]$ is the classical 
Euclidean action integral along the path $x(\tau)$ with $0 < \tau \leq \beta \hbar$.  In the following we will denote the Euclidean action as $\mathcal{A}$.
The integrand in $\mathcal{A}$, i.e., $(M/2) \dot{x}^2(\tau) + V(x(\tau))$, can be
identified with the classical Hamiltonian function, in which the
momentum $p$ is substituted for $M \dot{x}$. One can also regard
(\ref{PIx}) as an expectation value of the functional $\exp[
-\int_0^{\beta\hbar} \! d\tau  \ \! V(x(\tau))/\hbar ]$ over the
(driftless) Brownian motion with the diffusion coefficient
$M/2\hbar$, and duration $\beta\hbar$, that starts at point $x_a$,
and terminates at $x_b$. The latter stochastic process is also known
as Brownian bridge.

\section{Local-time representation of path integrals: heuristic approach}
\label{sec:Heuristic}

The purpose of this section is twofold. Firstly, we would like to motivate a
need for reformulation of PI's 
in the language of local-time stochastic process. In particular, we point out when such 
a reformulation can be more pertinent than the  conventional ``sum over
histories'' prescription.    
Secondly, we wish to outline a heuristic construction of
the local-time representation of PI's. More rigorous and explicit (but less
intuitive) formulation of 
PI's over ensemble of local times will be presented in the subsequent Section. 

To  provide a physically sound  motivation for the local-time representation of PI's  we 
follow  
exposition of Paulin {\it et al.}
in Ref.~\cite{Paulin}. To this end we first consider the diagonal elements
of the Bloch density matrix, i.e.,  $\rho(x_a,x_a,\beta)$  (often referred to as the Boltzmann
density). Upon shifting $x \rightarrow x+x_a$, and setting $x =
\lambda \xi$, $\tau = s \beta \hbar$ ($\lambda \equiv \sqrt{\beta
\hbar^2 / M}$ is the thermal de~Broglie wavelength), the PI
(\ref{PIx}) can be reformulated in terms of dimensionless quantities
$s$ and $\xi(s)$ as
\begin{equation} \label{PIxResc}
\rho(x_a,x_a,\beta) \ = \
\frac{1}{\lambda} \int_{x(0)=0}^{x(1)=0} \!\!\mathcal{D}\xi(s) \ \! \exp \left\{-\int_{0}^{1} \!\!\!ds
\left[\frac{1}{2}\dot{\xi}^2 + \beta V(x_a + \lambda \xi) \right] \right\} .
\end{equation}
Note in particular, that in contrast to $\mathcal{D}x(\tau)$ the
measure $\mathcal{D}\xi(s)$ does not explicitly depend on $\beta$,
and thus $\beta$-dependent parts in the PI are under better control.
Such a rescaled representation is particularly useful when
discussing large- and/or small-$\beta$ behavior of the path integral
in question. Path fluctuations in the potential are controlled by 
$\lambda \propto \sqrt{\beta}$, and the factor $\beta$ quantifies
the significance of the potential $V$ with respect to the kinetic term.

\begin{figure}
\includegraphics[scale=0.65]{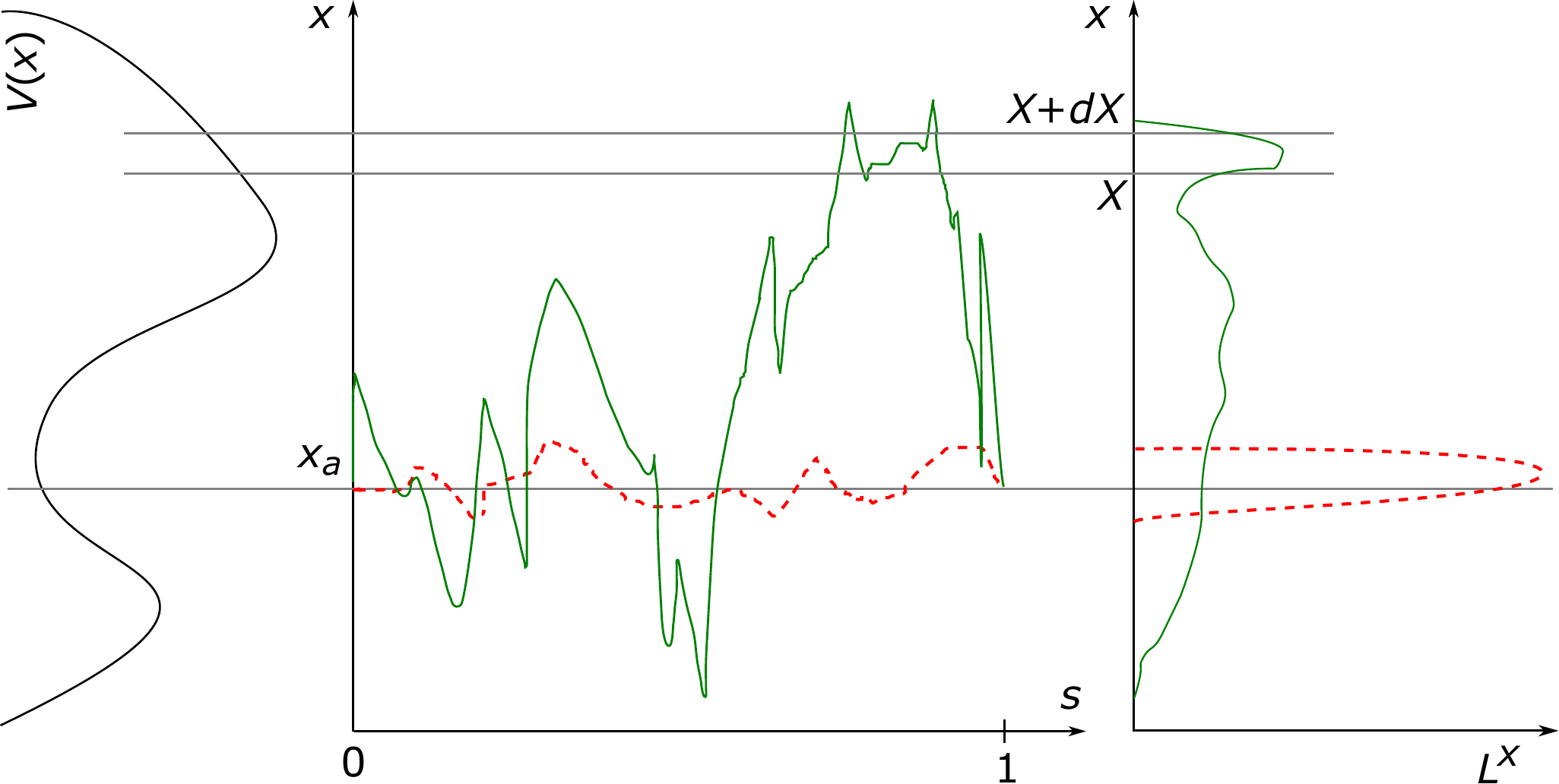}
\caption{In the middle, two typical paths $x(s\beta\hbar)=x_a+\lambda\xi(s)$ are plotted as functions of the
dimensionless time $s$. The solid green path, representing a typical trajectory with a high value of $\beta$, exhibits large fluctuations,
whereas the dashed red path, corresponding to small $\beta$, stays in the vicinity of the initial and final point $x_a$.
On the right, two local-time profiles $L(x)$ are shown. The broad one (solid green) arises from the violently fluctuating
path $x(s\beta\hbar)$, whereas the narrow one (dashed red) corresponds to the path with small fluctuations. On the left,
we depict a generic potential $V(x)$.}
\label{fig:paths}
\end{figure}

For small $\beta$ (i.e., high temperature), typical paths
$x(s\beta\hbar)=x_a+\lambda \xi(s)$ stay in the vicinity of the
point $x_a$, as depicted in Fig.~\ref{fig:paths},
and therefore a systematic Wigner--Kirkwood  expansion can be readily developed by
Taylor-expanding the potential part of the action~\cite{JizbaZatl}.

When $\beta$ is large (i.e., low temperature), the trajectories
$x(s\beta\hbar)$ fluctuate heavily around the value $x_a$, and
the potential term $V$ dominates over the kinetic one. From the statistical physics point of view, the most important contribution 
to the low temperature 
behavior of the path integral (\ref{PIxResc}) should come from those
paths that spend a sizable amount of time near the {\em global} minimum of the
potential $V(x)$.
For this reason, it is important to be able to keep track of the time which a
given path spends in an infinitesimal neighborhood of an
arbitrary point $x$.

Let us define, for each Wiener trajectory $x(\tau)$ present in the Feynman path integral (\ref{PIx}) the ensuing \emph{local time} as
\begin{equation} \label{LTDef}
L^X(\tau) \ = \ \int_0^{\tau}\! d\tau' \ \! \delta \!\left( X - x(\tau') \right) \,  .
\end{equation}
Since the local time $L^X(\tau)$ is a functional of the random trajectory $x(\tau')$ for $0 < \tau'< \tau$, it represents a random variable. 
From the definition (\ref{LTDef}) we can immediately see that 
$L^X \geq 0$ for all $X \in \mathbb{R}$, $\int_\mathbb{R} dX \ \! L^X(\tau)  = \tau$ and that $L^X$ has a compact support. In addition, 
it can be proved~\cite{RevuzYor,Marcus} that local-time trajectories $L^X$ are, with probability one, continuous curves which (similarly as trajectories in the underlying 
Wiener process) are nowhere differentiable.
In Fig.~\ref{fig:paths} we depict two examples of representative local-time trajectories.
An extensive mathematical discussion of properties of the local time can be found, e.g. in Refs.~\cite{Marcus,Borodin}.

With the definition (\ref{LTDef}) the potential part of the Euclidean action can be recast into form
$\int_\mathbb{R} dX \ \! L^X(\beta\hbar) V(X) $. A local-time representation of the Bloch density matrix $\rho(x_a,x_b,\beta)$ is then given by
\begin{equation} \label{HeurForm}
\rho(x_a,x_b,\beta) \ = \ \int \mathcal{D}L^x \ \! \mathcal{W}[L;\beta,x_a,x_b] \ \!\delta \! \left( \int_\mathbb{R} dX \ \! L^X  - \beta\hbar \right) \exp \left[-\frac{1}{\hbar} \int_\mathbb{R}  dX \ \! L^X V(X) \right] ,
\end{equation}
where the PI ``sum'' is taken over all local-time trajectories $L^x$, with $x$ being the independent variable (not to be mistaken with the Wiener trajectory $x(\tau)$).
The $\delta$-function enforces the normalization constraint mentioned above. Basically, transition to the
local-time description represents a change (or a functional substitution) of stochastic variables  $x(\tau) \rightarrow L^x(\beta\hbar)$. The weight  factor $\mathcal{W}$
appearing in (\ref{HeurForm}) can be formally written in the form  
\begin{eqnarray}
\mathcal{W}[L;\beta,x_a,x_b]  \ = \  \exp
\left[-\frac{1}{\hbar}\int_{0}^{\beta\hbar}\!\! d\tau \ \! \frac{M}{2}\dot{x}^2
\right] \ \! \det\left(\frac{\delta L^x(\beta\hbar)}{\delta x(\tau)}
\right)^{-1}\, .
\label{III.7.a}
\end{eqnarray}
It is a function of  $x_a$, $x_b$ (which are implicitly present in $L^x(\beta\hbar)$) and $\beta$, 
and a functional of the local time $L^x$.  
Of course, these cavalier manipulations do not have more than a heuristic nature, and
it is, indeed, a non-trivial task to determine $\mathcal{W}$ directly from (\ref{III.7.a}).  For this reason, we will in the following Section
tackle this problem indirectly.


\section{Local-time representation of path integrals: derivation}
\label{sec:LT}

In this Section, we present a derivation of the local-time representation of the Bloch density matrix (\ref{rhoDef}). 
Initially, we limit ourselves to considering the case of the diagonal part,
$x_b=x_a$, and arrive at the key result (\ref{DiagLapl}),
which expresses the matrix elements in the ensuing Laplace picture with respect to $\beta$. This intermediate outcome is shown
to agree with the Sturm--Liouville theory. In the next step, we generalize the latter result 
to the off-diagonal elements ($x_b \neq x_a$).
The inverse Laplace transform will then yield
the sought local-time representation of PI [cf. Eq.~(\ref{OffDiag})].

\subsection{Field-theoretic representation}
\label{sec:FieldTheor}

It follows from the definition (\ref{rhoDef}) that the function $\rho(x_a,x_b,\beta)$ satisfies the heat equation
\begin{equation} \label{DiffusionEq}
\left[ \frac{\partial}{\partial \beta} -\frac{\hbar^2}{2M} \frac{\partial^2}{\partial x_b^2} + 
V(x_b) \right] \rho(x_a,x_b,\beta) \ = \ 0\, ,
\end{equation}
with the initial condition $\rho(x_a,x_b,0_{+}) = \delta(x_a-x_b)$.
This is merely a Wick-rotated ($t \rightarrow -i \hbar \beta$) analogue of the
Schr\"{o}dinger equation. The Feynman--Kac
formula~\cite{Kac1949,Feynman1948,Friedlin:85} then ensures  that the PI
(\ref{PIx}) can be calculated by solving corresponding parabolic differential equation (\ref{DiffusionEq}).

In the Laplace picture Eq.~(\ref{DiffusionEq}) takes the form
\begin{equation} \label{LapDifEq}
\left[ E -\frac{\hbar^2}{2M} \frac{\partial^2}{\partial x_b^2} + V(x_b) \right] \widetilde{\rho}(x_a,x_b,E) \ = \ \delta(x_a-x_b)\,  ,
\end{equation}
with $\widetilde{\rho}(x_a,x_b,E) = \int_0^\infty d\beta e^{-\beta E} \rho(x_a,x_b,\beta)$. Eq.~(\ref{LapDifEq}) implies that $\widetilde{\rho}$ is nothing but the Green function of the operator $E+\hat{H}$. 
With the benefit of hindsight, we represent the Green function $\widetilde{\rho}$ as a path integral over fluctuating fields --- so-called functional integral~\cite{Ramond}. This is rather standard strategy in Quantum Field Theory~\cite{Zinn,BJV}, and in our case it yields
\begin{equation} \label{FieldRepre}
\widetilde{\rho}(x_a,x_b,E) \ = \
\frac
{\int_{\psi(X_{-})=0}^{\psi(X_{+})=0} \mathcal{D}\psi(x) \ \! \psi(x_a) \psi(x_b) \ \! e^{ -\frac{1}{2} \mathcal{A}^E[\psi] } }
{\int_{\psi(X_{-})=0}^{\psi(X_{+})=0} \mathcal{D}\psi(x) \ \!
e^{ -\frac{1}{2} \mathcal{A}^E[\psi]  } }\,  ,
\end{equation}
where
\begin{align} \label{ActionHO}
\mathcal{A}^E[\psi]
&\ \equiv \ \int_{X_{-}}^{X_{+}} \!\!\!dx \ \! \psi(x) \left[-\frac{\hbar^2}{2M} \frac{d^2}{d x^2} + V(x) + E \right] \psi(x) \nonumber\\[2mm]
&\ = \ \int_{X_{-}}^{X_{+}} \!\!\!dx  \left[ \frac{\hbar^2}{2M} \psi'(x)^2 + ( V(x) + E ) \psi(x)^2 \right],
\end{align}
is the Euclidean action functional of the field-theoretic path integral. The super-index ``$E$'' in $\mathcal{A}$ indicates the shift in the potential $V(x)$ by the amount 
$E$. 
Here, we have confined our quantum-mechanical system within a finite box
$[X_-,X_+]$, with $X_- \ll \min\{x_a,x_b\}$ and  $X_+ \gg \max\{x_a,x_b\}$. A
real scalar field $\psi(x)$ satisfies Dirichlet boundary conditions $\psi(X_-) =
\psi(X_+) = 0$ so as to ensure the validity of the operations being performed. 


\subsection{Replica trick}
\label{sec:Replica}

Since we will ultimately want to invert the Laplace transform to regain from
$\widetilde{\rho}(x_a,x_b,E)$ the original Bloch density matrix
$\rho(x_a,x_b,\beta)$, we cannot treat the denominator in
(\ref{FieldRepre}) as an irrelevant normalization constant (which is the usual practice in
Quantum Field Theory) but we have to take care of its $E$-dependence. To this end we take advantage 
of the formula
\begin{equation}
\frac{a}{b} \ = \ \lim_{D \rightarrow 0} a b^{D-1}\,  ,
\label{Rep.1a}
\end{equation}
which is a simple version of the so-called {\it replica trick}. [The usual replica-trick formula (cf. e.g., Ref.~\cite{Parisi:87}) can be obtained from (\ref{Rep.1a}) by integrating both sides with respect to $b$ and subsequently dividing by $a$.] With the help of (\ref{Rep.1a}) we can rewrite (\ref{FieldRepre}) as a multidimensional functional integral
\begin{equation} \label{AfterRepl}
\widetilde{\rho}(x_a,x_b,E) \ = \ \lim_{D \rightarrow 0} \frac{2}{D} \int_{\bm{\psi}(X_{-})=\bm{0}}^{\bm{\psi}(X_{+})=\bm{0}}
\!\! \mathcal{D}\bm{\psi}(x) \ \! \bm{\psi}(x_a) \cdot \bm{\psi}(x_b) \ \!
e^{- \sum_{\sigma=1}^D \mathcal{A}^E[\psi_\sigma] }\,  ,
\end{equation}
where the multiplet ${\bm \psi} = (\psi_1,\ldots,\psi_D)$ is a $D$-component ``replica" field in $1+0$ dimensions, $\bm{\psi}(x_a) \cdot \bm{\psi}(x_b)$ denotes the scalar product $\sum_{\sigma=1}^D \psi_\sigma(x_a) \psi_\sigma(x_b)$, and we have rescaled the fields by a factor of $\sqrt{2}$ in passing.
The factor $1/D$ results from a PI generalization of the well know mean-value identity $\langle x_i\ \! y_i \rangle = \langle {\bf{x} }\cdot {\bf{y}}\rangle/D$ valid for any two vectors in $D$-dimensional statistically isotropic environments.

As a side remark, note that we may now invert the Laplace transform, using the trivial identity
\begin{equation} \label{InverseLap}
\int_0^\infty d\beta \ \! e^{-\beta E} \delta(\beta-c) \ = \ e^{-c E}\;\;\;\,\,\,\, \mbox{for} \;\;\;\; c \ > \ 0\, ,
\end{equation}
to obtain the representation
\begin{equation} \label{ReplRepre}
\rho(x_a,x_b,\beta) \ = \ \lim_{D \rightarrow 0} \frac{2}{D}
\int_{\bm{\psi}(X_{-})=\bm{0}}^{\bm{\psi}(X_{+})=\bm{0}} \!\! \mathcal{D}\bm{\psi} \ \! \ \! \bm{\psi}(x_a) \cdot \bm{\psi}(x_b)
\ \!\ \! \delta\! \left( \int_{X_-}^{X_+} \!\!\! \bm{\psi}(x)^2 dx - \beta \right)
e^{- \sum_{\sigma=1}^D \mathcal{A}^{E=0}[\psi_\sigma] }\, .
\end{equation}
Upon scaling $\bm{\psi} \rightarrow \sqrt{\beta} \bm{\psi}$, this agrees with the formula (2.9) in Ref.~\cite{Luttinger1982}. Though the result (\ref{ReplRepre}) generalizes to arbitrary number of dimensions of the $x$-space, i.e., $x \in \mathbb{R}^d$, our further development will be illustrated (for simplicity's sake) only on the one-dimensional case.

\subsection{Connection with radial harmonic oscillator}
\label{sec:Radial}

In order to derive the weight factor $\mathcal{W}$ (cf. Eq.~(\ref{III.7.a}))  we have arrived at the representation (\ref{ReplRepre}) with $D$ 
replica fields.  This form is still not very transparent, and a further simplification step is needed to get rid of an explicit dependence of the measure on $D$. To this end, we note that $\sum_{\sigma=1}^D \mathcal{A}^E[\psi]$ is in fact the action of a $D$-dimensional harmonic oscillator with the ``time'' variable $x$, ``position'' variable $\psi$, mass $\hbar^2/M$, and time-dependent ``frequency'' $V(x) + E$. Considering for a moment only diagonal matrix elements, $x_b=x_a$, spherical symmetry in the replica field space allows to reduce the path integral (\ref{AfterRepl}) to its radial part. Due to the boundary conditions, $\bm{\psi}(X_{-})=\bm{\psi}(X_{+})=\bm{0}$, only the zero-angular-momentum ($s$-wave) contribution is non-vanishing (generally a weighted sum over radial PI's with different angular momenta would be required). This will be rigorously justified at the end of this Section. The corresponding {\it radial} PI representation for (\ref{AfterRepl}) reads~\cite{Grosche,Zinn,Kleinert} 
\begin{equation} \label{DiagLapl}
\widetilde{\rho}(x_a,x_a,E) \ = \
\lim_{D \rightarrow 0} \frac{2}{D \Omega(D)} \lim_{\eta_\pm \rightarrow 0} (\eta_- \eta_+)^{\frac{1-D}{2}}
\int_{\eta(X_-)=\eta_-}^{\eta(X_+)=\eta_+} \!\! \mathcal{D}\eta(x) \ \! \eta^2(x_a) \ \!
e^{ -A_D^E[\eta] }\, .
\end{equation}
%
%
Here, the radial part $\eta \equiv \sqrt{\bm{\psi}^2}$ of the $D$-dimensional replica field $\bm{\psi}$ is always non-negative, i.e., $\eta(x) \geq 0$; the area of a unit sphere in $D$ dimensions, $\Omega(D) = 2 \pi^{D/2} / \Gamma(D/2)$, may be replaced by its small-$D$ asymptotic form $\Omega(D) \sim D$; and $\eta_\pm$ have been introduced to regularize the origin of the $\bm{\psi}$-space. The new action functional 
\begin{equation}
\mathcal{A}_D^E[\eta] \ \equiv \ \mathcal{A}^E[\eta] \ + \  \int_{X_-}^{X_+} \!\! dx \ \! \frac{M}{\hbar^2} \frac{(D\!-\!1)(D\!-\!3)}{8 \eta^2(x)} \, ,
\end{equation}
is the Euclidean action functional of the  \emph{radial harmonic oscillator}~\cite{Kleinert,Grosche,Steiner}. It contains an additional centrifugal potential term (Edwards--Gulyaev or Langer term~\cite{E-G-64,Kleinert,Inomata:66}), which emerges from Bessel function $I_{D/2-1}$ present in the finite sliced form of the radial PI (\ref{DiagLapl}). At this point we should stress that in contrast to the quantum-mechanical radial PI, one can use safely the 
asymptotic expansion for the Bessel function $I_{D/2-1}$ (see, e.g. Ref.~\cite{Gradshteyn}):
\begin{eqnarray}
&&I_{\mu}(y_j) \ \sim \ \frac{1}{\sqrt{2\pi y_j}} \ \! e^{y_j-(\mu^2 -1/4)/2y_j}, \;\;\;\;\;\;\; \;\;(|y_j| \gg 1, \; \mbox{Re}[y_j] > 0)\, ,\nonumber \\[2mm]
&&y_j \ = \ (m/\varepsilon \hbar)r_jr_{{j-1}}\, ,
\label{asymptot-b}
\end{eqnarray}
in the Euclidean PI sliced form. Here the infinitesimal ``time'' slice $\varepsilon$ is related to the number of slices $N$ via the relation 
$\varepsilon = \hbar \beta/N$. In quantum mechanic this is a problematic step because (\ref{asymptot-b}) requires  $\mbox{Re}[y_j] > 0$ while there
$\mbox{Re}[y_j] = \mbox{Re}[(m/i\varepsilon \hbar)r_jr_{{j-1}}] = 0$. 

Fortunately, the PI for radial harmonic oscillator is exactly solvable even in the case of $x$-dependent oscillator frequency. The solution reads~\cite{Grosche}
\begin{eqnarray} 
(\eta_2 x_2 | \eta_1 x_1 )_{D} \ & \equiv &
\int_{\eta(x_1)=\eta_1}^{\eta(x_2)=\eta_2} \!\!\!\mathcal{D}\eta(x)
\exp\left\{ -\int_{x_1}^{x_2} \!\!\! dx \left[\frac{\hbar^2}{2 M}\eta'^2 + (V(x)+E)\eta^2 + \frac{M}{\hbar^2} \frac{(D\!-\!1)(D\!-\!3)}{8 \eta^2} \right] \right\} \nonumber \\[2mm]
&=& \ \frac{\hbar^2}{M} \frac{\sqrt{\eta_1 \eta_2}}{G(x_1)} I_{D/2-1}\left(\frac{\hbar^2}{M} \frac{\eta_1 \eta_2}{G(x_1)}\right)
\exp\left[- \frac{\hbar^2}{2 M} \left(\frac{F'(x_2)}{F(x_2)} \eta_2^2 - \frac{G'(x_1)}{G(x_1)} \eta_1^2 \right) \right].
\label{RadialHOSol}
\end{eqnarray}
The functions $F(x)$ and $G(x)$ are two independent solutions of the differential equation
\begin{equation} \label{DiffEqFG}
\left[\hat{H}+E\right]y(x) \ = \ \left[ -\frac{\hbar^2}{2 M} \frac{d^2}{dx^2} + V(x) + E \right] y(x) \ = \ 0\, ,
\end{equation}
with the initial conditions $F(x_1)=0$ and $F'(x_1) = 1$, and $G(x_2)=0$ and $G'(x_2) = -1$. In addition, the Wronskian $W(F,G) \equiv F(x) G'(x) - F'(x) G(x)$ is independent of $x$, as can be proved by differentiation and by using the fact that $F$ and $G$ both satisfy Eq.~(\ref{DiffEqFG}). By equating the values of $W(F,G)$ at points $x_1$ and $x_2$, and taking into account the initial conditions for $F$ and $G$, we find a useful identity $F(x_2)=G(x_1)$.

Now, the PI in (\ref{DiagLapl}) can be sliced at point $x_a$, and expressed as 
\begin{eqnarray}
\widetilde{\rho}(x_a,x_a,E)  \ = \ \int_0^\infty \! d\eta_a \ \!(\eta_+ X_+ | \eta_a x_a )_{D}\ \! \eta_a^2 \ \! (\eta_a x_a | \eta_- X_- )_{D}\, .
\end{eqnarray}
The limits in Eq.~(\ref{DiagLapl}) are readily carried out with the help of the asymptotic formulas $I_{D/2-1}(z)~\approx~(z/2)^{D/2-1} / \Gamma(D/2)$, and $\Gamma(z)~\approx~1/z$, valid for $z \rightarrow 0_+$. Subsequent integration over $\eta_a$  brings (\ref{DiagLapl}) to form
\begin{equation} \label{DiagGreenExpl}
\widetilde{\rho}(x_a,x_a,E) \ = \
- \frac{2M}{\hbar^2} \frac{F_1(x_a) G_2(x_a)}{F_1(x_a) G'_2(x_a) - F'_1(x_a) G_2(x_a)}\,  ,
\end{equation}
where $F_1(x)$ solves Eq.~(\ref{DiffEqFG}) with initial conditions
$F_1(X_{-}) = 0$ and $F_1'(X_{-}) = 1$, and $G_2(x)$ solves the same equation with $G_2(X_{+}) = 0$ and $G_2'(X_{+}) = -1$. The denominator in (\ref{DiagGreenExpl}) is the Wronskian $W(F_1,G_2)$.  The full derivation is given in Appendix~\ref{sec:App}.

Although rather explicit, Eq.~(\ref{DiagGreenExpl}) is not well suited for the Laplace transform inversion, since functions $F(x)$ and $G(x)$ contain $E$ in a non-trivial way, which, in addition, significantly hinges on the actual form of  $V(x)$. For formal manipulations it is still better the employ the PI representation (\ref{DiagLapl}). 
For instance, using Eq.~(\ref{InverseLap}) we can easily invert 
the Laplace transform to return from $E$ back to the $\beta$-variable, namely
\begin{eqnarray}
\mbox{\hspace{-1mm}}\rho(x_a,x_a,\beta)  = 
\lim_{D \rightarrow 0} \frac{2}{D^2} \lim_{\eta_\pm \rightarrow 0} (\eta_- \eta_+)^{\frac{1-D}{2}}
\!\!\int_{\eta(X_-)=\eta_-}^{\eta(X_+)=\eta_+} \!\!\! \mathcal{D}\eta(x) \ \! \eta^2(x_a) \ \!
 \delta \!\left( \int_{X_-}^{X_+} \!\!\!\!\! \eta^2 dx - \beta \right)\!
e^{ -A_D^{E=0}[\eta] }\, .
\label{Boltzmann.II}
\end{eqnarray}
Note that we have utilized the asymptotic form $\Omega(D) \sim D$ which holds for $D \ll 1$.
We shall see shortly that (\ref{Boltzmann.II}) can be straightforwardly related to the local-time PI representation of the Boltzmann density matrix.

Let us finally comment on the higher-angular-momentum terms which, as claimed, should not contribute to the expression (\ref{DiagLapl}). For arbitrary angular momentum $\ell \geq 0$, we employ formula (\ref{RadialHOSol}) with a slight modification $D \rightarrow D+2\ell$. 
Now, for example, in the limit $\eta_- \rightarrow 0$, this goes like $(\eta_a x_a | \eta_- X_- )_{D+2\ell} \propto \eta_-^{\ell+D/2-1/2}$, which, multiplied by the prefactor $\eta_-^{1/2-D/2}$, implies the behavior $\sim \eta_-^\ell$. That is, only the ($\ell=0$)-term can give a non-vanishing contribution.

\subsection{Connection with the Sturm--Liouville problem}
\label{sec:Sturm}

Consider again Eq.~(\ref{LapDifEq}) and a finite interval $x \in [X_-,X_+]$. The corresponding Green 
function of the operator $\hat{H}+E$ can be easily constructed (at least formally) with the help of the 
Sturm--Liouville theory~\cite{Vladimirov,Zettl}. An immediate consequence
of the latter is that for $x_a<x_b$ the Green function has the form
\begin{equation} \label{SturmGreen}
\widetilde{\rho}(x_a,x_b,E) \  = \
- \frac{2M}{\hbar^2} \frac{F(x_a) G(x_b)}{W(F,G)}\,  ,
\end{equation}
where the functions $F(x)$ and $G(x)$ satisfy Eq.~(\ref{DiffEqFG}) with the initial conditions $F(X_-)=0$ and $F'(X_-) = 1$, and $G(X_+)=0$ and $G'(X_+) = -1$, respectively. In addition, the above Green function should be  symmetric due to the Hermitian nature of $\hat{H}$. 

The Sturm--Liouville theory ensures that the solution to the second-order differential equation (\ref{DiffEqFG}) is unique, when specifying the values of $y(x_0)$ and $y'(x_0)$ at some point $x_0$. Therefore, the functions $F$ and $G$ must coincide with $F_1$ and $G_2$ of Eq.~(\ref{DiagGreenExpl}), and the diagonal part of (\ref{SturmGreen}), i.e. $\widetilde{\rho}(x_a,x_a,E)$, reduces to expression (\ref{DiagGreenExpl}). This is an important consistency check of our representation (\ref{DiagLapl}).

\subsection{Extension to off-diagonal matrix elements}
\label{sec:OffDiag}

Let us now generalize the PI representation  (\ref{DiagLapl}) to the full Bloch density matrix, i.e., we wish to include also the off-diagonal matrix elements, $x_b \neq x_a$. If we go back to the replica representation (\ref{AfterRepl}), we realize that the requirement $x_b \neq x_a$ spoils rotational symmetry in the replica field space, and thus precludes straightforward reduction to a radial path integral. Instead of refining the reduction procedure, we simply make a guess, which, as we prove in Appendix~\ref{sec:App}, coincides with the well-established Sturm--Liouville Formula (\ref{SturmGreen}). Our guess is based on mathematical results presented in~\cite{Borodin}. In particular, we claim that
the extension of the representation (\ref{DiagLapl}) to off-diagonal matrix elements should read
\begin{equation} \label{LaplOffDiag}
\widetilde{\rho}(x_a,x_b,E)
= \lim_{D \rightarrow 0} \frac{2}{D^2} \lim_{\eta_\pm \rightarrow 0} (\eta_- \eta_+)^{\frac{1-D}{2}}
 \int_{\eta(X_-)=\eta_-}^{\eta(X_+)=\eta_+} \!\!\! \mathcal{D}\eta(x) \ \! \eta(x_a) \eta(x_b) \ \!
e^{ -\mathcal{A}^E_\Delta[\eta] }\, ,
\end{equation}
with the action functional
\begin{equation}
\mathcal{A}^E_\Delta[\eta] \ \equiv \ \mathcal{A}^E[\eta] \ + \ \int_{X_-}^{X_+} \!\! dx \ \! \frac{M}{\hbar^2} \frac{\Delta(x)}{8 \eta^2(x)}\,  ,
\end{equation}
where $\mathcal{A}^E[\eta]$ is defined in Eq.~(\ref{ActionHO}), and
$\Delta(x)$ is a piecewise constant function
\begin{equation}
\Delta(x) =
\begin{cases}
  -1 & {\rm for~} x \in [x_a , x_b] \\
  (D-1)(D-3) & {\rm otherwise} .
\end{cases}
\end{equation}

At this point we can invert the Laplace transform  with the help of Eq.~(\ref{InverseLap}). 
As a result, we obtain the sought local-time PI representation of the Bloch density matrix (\ref{rhoDef}), namely
\begin{equation} \label{OffDiag}
\rho(x_a,x_b,\beta)
= \lim_{D \rightarrow 0} \frac{2}{D^2} \lim_{\eta_\pm \rightarrow 0} (\eta_- \eta_+)^{\frac{1-D}{2}}
 \int_{\eta(X_-)=\eta_-}^{\eta(X_+)=\eta_+} \!\!\! \mathcal{D}\eta \ \! \ \!\eta(x_a) \eta(x_b) \ \!
 \delta \!\left( \int_{X_-}^{X_+} \!\!\!\!\! \eta^2 dx - \beta \right)
e^{ - \mathcal{A}^{E=0}_\Delta[\eta] }\, .
\end{equation}
Here, integrations over $\eta(x)$ run from $0$ to $+\infty$, i.e., the paths $\eta(x)$ are non-negative.
Comparing this result with the anticipated heuristic form (\ref{HeurForm}), we
can identify $\eta^2(x) = L^X(\beta\hbar)/\hbar$. 
Representation (\ref{OffDiag}) allows us to identify the weight factor (\ref{III.7.a}) with
\begin{equation} \label{Weights}
\mathcal{W}_D[\eta;\beta,x_a,x_b] \ = \
\frac{2}{D^2} (\eta_- \eta_+)^{\frac{1-D}{2}}
\eta(x_a) \eta(x_b)
\exp \left\{ -\int_{X_-}^{X_+} \!\!\! dx \left[\frac{\hbar^2}{2 M}{\eta'}^2 + \frac{M}{\hbar^2}\frac{\Delta(x)}{8 \eta^2} \right] \right\} .
\end{equation}
Contrary to expectation, the right-hand-side  of this expression does not depend on $\beta$. Sub-index $D$ in $\mathcal{W}_D$ indicates that the weight factor must be regularized when we pull it out of the PI (\ref{OffDiag}). 
By analogy with quantum mechanics one can represent (\ref{OffDiag}) in the discretized  time-sliced form. In such a case the weight  $\mathcal{W}_D$ would be a product of terms involving the Bessel functions $I_{D/2-1}$, if $\Delta(x)=(D-1)(D-3)$, or $I_0$, if $\Delta(x)=-1$ (see Ref.~\cite{Grosche}).

Last but not least, expressions (\ref{OffDiag})-(\ref{Weights}) indicate that the square root of $L^X$ is (at least from physicist's point of view) more convenient variable to describe local-time trajectories than $L^X$ alone.


\section{Functionals of the local time}
\label{sec:Functionals}

Formula (\ref{OffDiag}) provides a way of rewriting the PI (\ref{PIx}) in terms of the local time. In this Section, we consider more general scenario, in which the initial path integral is of the form
\begin{equation} \label{PIFunc}
\bar{F}(x_a,x_b,\beta) \ \equiv \
\int_{x(0)=x_a}^{x(\beta\hbar)=x_b} \!\!\!\mathcal{D}x(\tau) \ \!
F[L] \ \!
\exp\left\{-\frac{1}{\hbar} \int_{0}^{\beta\hbar} \!\!\!d\tau \left[\frac{M}{2}\dot{x}^2 + V(x) \right] \right\} ,
\end{equation}
where $F$ is an arbitrary functional of the local time $L^X(\beta \hbar)$, which itself is (as seen in Section~\ref{sec:Heuristic}) a functional of the paths $x(\tau)$. 
Relation (\ref{PIFunc}) represents a local-time analog of the Feynman--Matthews--Salam formula~\cite{LvW,BJV}. 

To bring it into more manageable form, we may observe that for any $X$, the action of $L^X$ in the PI  (\ref{PIFunc}) can be taken over by the functional derivative $-\hbar \delta/\delta V(X)$, acting on the exponential. This becomes transparent after rewriting the potential part as $\int_0^{\beta\hbar} \! d\tau \ \! V(x(\tau)) = \int_{\mathbb{R}} 
dX \ \! V(X) L(X)$.  The entire functional $F[L]$ can be therefore pulled out of the path integral, which then allows to write
\begin{equation}
\bar{F}(x_a,x_b,\beta) \ = \
F \left[-\hbar  \frac{\delta}{\delta V} \right]
\rho(x_a,x_b,\beta) \, .
\end{equation}
When we employ the local-time representation of PI for $\rho(x_a,x_b,\beta)$ (cf. Eq.~(\ref{OffDiag})), each functional derivative
$-\hbar \delta/\delta V(x)$ will produce the term $\hbar \eta^2(x)$. In such a
way $\bar{F}$ can be written as 
\begin{eqnarray}
&&\mbox{\hspace{-11mm}}\bar{F}(x_a,x_b,\beta)\nonumber \\
&&\mbox{\hspace{-11mm}}= \ \lim_{D \rightarrow 0} \frac{2}{D^2} \lim_{\eta_\pm \rightarrow 0} (\eta_- \eta_+)^{\frac{1-D}{2}}
 \int_{\eta(X_-)=\eta_-}^{\eta(X_+)=\eta_+} \!\!\!\!\! \mathcal{D}\eta \ \! \ \!  \eta(x_a) \eta(x_b) \ \!\delta \!\left( \int_{X_-}^{X_+} \!\!\!\!\! \eta^2 dx - \beta \right)
F[\hbar \eta^2]
e^{ - \mathcal{A}^{E=0}_\Delta[\eta] }\, ,
\label{LTFunc}
\end{eqnarray}
where, strictly speaking, the functional $F[\cdots]$ is regularized in such a way that it depends on $L^X$ only for $X \in [X_-,X_+]$, and $X_\pm$ are sent to $\pm\infty$ only at the end of the calculation.

First, let us make the simple observation that the formula (\ref{LTFunc})  reduces to (\ref{OffDiag}) for the choice $F[L] = 1$.
One of the most important mean values of a local-time functional, as evaluated with Eq.~(\ref{LTFunc}), is the mean of $\exp(-\int_{\mathbb{R}} {dX L^X j(X)})$ which gives the moment-generating functional. The local-time moment structure is particularly pertinent in various perturbative expansions, including low- and high-temperature expansions (see Section~\ref{sec:Asymp}). Another important example, namely the case of a one-point distribution function will be discussed in the following subsection.

In passing we should note, that should we have started from (\ref{ReplRepre}) and repeated the above procedure, an analog of Eq.~(\ref{LTFunc}) for higher-dimensional spaces, $x \in \mathbb{R}^d$, could be easily obtained. This would include the $D$-dimensional replica 
field $\bm{\psi}$ in the $d$-dimensional Euclidean configuration space.
\subsection{Example: One-point distribution function at the origin}
\label{sec:Example}

Simple, though quite important consequence of Eq.~(\ref{LTFunc}) is that it readily provides the $N$-point distribution functions of the local time. This is achieved when 
we set $F[L] = \prod_{n=1}^N \delta (L^{X_n} - L_n )$. In order to see what is involved let us now illustrate
the calculation for $N=1$ (with $L_1 \equiv L$). Our discussion will be greatly simplified by considering only a free particle (i.e., $V(x) = 0$) that starts and ends at the origin, i.e., $x_a = x_b = 0$. This corresponds to a stochastic process known as Brownian bridge. Our goal is to derive  the one-point distribution function, denoted $p(L;\beta)$, of the local time at $X=0$. We define $p(L;\beta)$ by Eq.~(\ref{PIFunc}) with $F[L]=\delta ( L^0 - L )$, and calculate it from the representation  (\ref{LTFunc}) as follows.

In the Laplace picture, $\widetilde{p}(L;E) = \int_0^\infty \! d\beta e^{-\beta E} p(L;\beta)$, the path integral (\ref{LTFunc}) can be sliced at $x_a=x_b=0$ so that
\begin{equation}
\widetilde{p}(L;E) =
\lim_{D \rightarrow 0} \frac{2}{D^2} \lim_{\eta_\pm \rightarrow 0} (\eta_- \eta_+)^{\frac{1-D}{2}}
\int_0^\infty \!\!\! d\eta_0 \ \! \eta_0^2  \ \! \delta(\hbar \eta_0^2 - L)
(\eta_+ X_+ | \eta_0~ 0)_D (\eta_0~ 0 | \eta_- X_-)_D\, .
\end{equation}
%
The $\eta_0$-integration can be done easily by realizing that  $\delta(\hbar \eta_0^2 - L) = \delta(\eta_0-\sqrt{L/\hbar})/2\sqrt{\hbar L}$. Furthermore, the limits in $\eta_\pm$ and $D$ can be carried out with the help of formulas (\ref{etaLimMinus}) and (\ref{etaLimPlus}) from Appendix~\ref{sec:App}. Consequently, we obtain
\begin{equation} \label{Ex1}
\widetilde{p}(L;E) =
\exp \left[ -\frac{L \hbar^2}{2M} \left( \frac{F'_1(0)}{F_1(0)} - \frac{G'_3(0)}{G_3(0)} \right) \right] ,
\end{equation}
where, for the free-particle case, $F_1(x) = \sinh[\sqrt{2ME/\hbar^2}(x-X_-)]/\sqrt{2ME/\hbar^2}$, and $G_3(x) = \sinh[\sqrt{2ME/\hbar^2}(X_+-x)]/\sqrt{2ME/\hbar^2}$, as one can straightforwardly verify. In the limit $X_\pm \rightarrow \pm\infty$, Eq.~(\ref{Ex1}) reduces to
\begin{equation}
\widetilde{p}(L;E) = e^{- \sqrt{2\hbar^2 E /M} L} ,
\end{equation}
and its inverse-Laplace transform yields
\begin{equation} \label{LTfree}
p(L;\beta) =
\frac{L \exp\left(- \frac{L^2 \hbar^2}{2\beta M} \right)}{\sqrt{2 \pi M \beta^3/\hbar^2}}\,  .
\end{equation}

We stress that $p(L;\beta)$ thus obtained is, in fact, the (unnormalized ) {\it joint} probability density for stochastic events ``$x(0) = 0 \rightsquigarrow x(\beta\hbar) = 0$" and ``$L^{0} = L$". By Bayes' theorem of the probability calculus, the desired conditional probability density $p[L^0 \!=\! L|x(0)\! =\! 0\! \rightsquigarrow \! x(\beta\hbar) \!=\! 0]$
is obtained from (\ref{LTfree}) by dividing $p(L;\beta)$ by the Brownian-bridge probability density $p[x(0)\! =\! 0\! \rightsquigarrow \! x(\beta\hbar) \!=\! 0]$, which is (omitting again normalization) $(2\pi \beta \hbar^2 / M)^{-1/2}$ (see, e.g., Ref.~\cite{FeynmanHibbs}).
Normalization factors mutually cancel in the fraction and we arrive at
\begin{eqnarray}
 p[L^0 \!=\! L|x(0)\! =\! 0\! \rightsquigarrow \! x(\beta\hbar) \!=\! 0] \ = \  \frac{\hbar^2 L \exp\left(- \frac{L^2 \hbar^2}{2\beta M} \right)}{{\beta M}}\,  ,
\end{eqnarray}
which is clearly normalized to $1$.
%
One can proceed along the same lines also in more complicated higher-dimensional
($N>1$)  cases.
Our result agrees with the one found through other means in Ref.~\cite{Borodin}.

\section{Asymptotic behavior of the Bloch density matrix}
\label{sec:Asymp}


A compelling feature of the local-time representation (\ref{OffDiag}) is that it naturally captures both small- and large-$\beta$ asymptotic regimes. This should be compared with the Feynman--Kac PI representation (\ref{PIx}), which is typically suitable only for the small-$\beta$ (i.e., large-temperature) analysis. The latter is epitomized either by WKB approximation~\cite{Zinn,Kleinert} or Wigner--Kirkwood expansion~\cite{JizbaZatl}. In the large-$\beta$ (small-temperature) limit, the spectral representation of the Gibbs operator, $e^{-\beta \hat{H}} = \sum_n e^{-\beta E_n} \ket{\phi_n} \bra{\phi_n}$, reduces the Bloch density matrix to the ground-state contribution
\begin{equation} \label{SpectralAsymp}
\rho(x_a,x_b,\beta) \ \overset{\beta \rightarrow \infty}{\sim}\ 
e^{-\beta E_{gs}}  \ \! \psi_{gs}^{*}(x_a) \psi_{gs}(x_b)\, ,
\end{equation}
that is not evident from the Feynman--Kac PI representation~\cite{Note-1}. In connection with Eq.~(\ref{SpectralAsymp}) it is useful to remind that
in $d=1$ the discrete bound states can all be chosen to be real~\cite{Messiah}, so that the Bloch density matrix is real and symmetric and can be written in the form
\begin{equation} \label{SpectralAsymp2}
\rho(x_a,x_b,\beta) \ = \ \sum_{n=0} e^{-\beta E_n } \psi_{n}(x_a) \psi_{n}(x_b) \ \overset{\beta \rightarrow \infty}{\sim}\ 
e^{-\beta E_{gs}}  \ \! \psi_{gs}(x_a) \psi_{gs}(x_b)\, .
\end{equation}

Let us first comment on the small-$\beta$ regime of the local-time representation, assuming $x_b=x_a$ for simplicity.  This case was discussed in detail in our previous article~\cite{JizbaZatl}. There one should first  Taylor-expand the potential $V(x)$ around the point $x_a$, and then expand the exponential part containing the structure $\int \eta^2(x)O(x-x_a) dx$, where
\begin{eqnarray}
O(x-x_a) \ = \ \beta \sum_{m\neq 0} \frac{V^{(m)}(x_a)}{m!} [\lambda (x-x_a)]^m\, .
\end{eqnarray}
After the term $e^{-\beta V(x_a)}/\lambda$  is factored out of the integral, the  individual summands of the ensuing series are of the form (\ref{LTFunc}) with the potential $V(x) = 0$, and functional $F[L] \propto \prod_n L^{x_n}/\hbar$. The latter can be related to the power expansion in $\beta$ presented in \cite{JizbaZatl} through the equality of representations (\ref{PIFunc}) and (\ref{LTFunc}). The whole Bloch density matrix  (containing also off-diagonal elements) can be treated similarly  in a full analogy with Ref.~\cite{JizbaZatl}.

The large-$\beta$ expansion of Eq.~(\ref{OffDiag}) can be conveniently studied after rescaling $\eta \rightarrow \sqrt{\beta}\eta$, in which case we can write
\begin{eqnarray} \label{OffDiagRescaled}
\rho(x_a,x_b,\beta)
&=& \ \lim_{D \rightarrow 0} \frac{2}{D^2} \lim_{\eta_\pm \rightarrow 0} (\eta_- \eta_+)^{\frac{1-D}{2}}
\int_{\eta(X_-)=\eta_-}^{\eta(X_+)=\eta_+}  \mathcal{D}\eta(x) \ \! \ \! \eta(x_a) \eta(x_b)   \nonumber\\[1mm]
&\times& \ \delta \! \left( \int_{X_-}^{X_+} \!\!\! \eta^2 dx - 1 \right)
\exp \left\{ -\int_{X_-}^{X_+} \!\!\! dx \left[\frac{\beta \hbar^2}{2 M}{\eta'}^2 + \beta V(x)\eta^2 + \frac{M}{\beta \hbar^2}\frac{\Delta(x)}{8 \eta^2} \right] \right\} \nonumber \\[3mm]
&=& \ \lim_{D \rightarrow 0} \frac{2}{D^2} \lim_{\eta_\pm \rightarrow 0} (\eta_- \eta_+)^{\frac{1-D}{2}}  \frac{\beta \delta^2}{\delta J(x_a) \delta J(x_b)}
\int_{c-i\infty}^{c+i\infty} \frac{d \kappa }{2\pi i} \int_{\eta(X_-)=\eta_-}^{\eta(X_+)=\eta_+} \mathcal{D}\eta(x)\nonumber\\[1mm]
&\times& \ \exp\left\{ - \beta \left[\int_{X_-}^{X_+} \!\!\! dx \left(\frac{\hbar^2}{2 M}{\eta'}^2 + V(x)\eta^2 - \kappa \eta^2\right)        
+ \kappa \right] \right\}\nonumber\\[1mm] 
&\times& \ \left. \exp\left\{ - \int_{X_-}^{X_+} \!\!\! dx \left[ \frac{M}{\beta \hbar^2}\frac{\Delta(x)}{8 \eta^2} + J\eta  \right] \right\}\right|_{J=0} ,
\end{eqnarray}
where $c$ is an arbitrary real number.
With the {\it method of images}~\cite{Kleinert,Grosche,Schulman:97} we can rewrite the radial PI involved as a superposition of two genuine one-dimensional PI's~\cite{Steiner} 
\begin{eqnarray}
&&\mbox{\hspace{0mm}}\int_{\eta(X_-)=\eta_-}^{\eta(X_+)=\eta_+}  \mathcal{D}^{R}\eta(x) \exp\left\{-\beta\left[\langle \eta| \hat{H}|
\eta \rangle -\kappa\left(\langle \eta|\eta \rangle -1 \right)  \right] - \langle J|\eta \rangle \right\}
\exp\left[  - \int_{X_-}^{X_+}  dx\ \! \frac{M}{\beta \hbar^2}\frac{\Delta(x)}{8 \eta^2}\right]_{\!R}\nonumber \\[3mm]
&&\mbox{\hspace{10mm}}= \ \int_{\eta(X_-)=\eta_-}^{\eta(X_+)=\eta_+}  \mathcal{D}\eta(x) \exp\left\{-\beta\left[\langle \eta| \hat{H}| \eta \rangle -\kappa\left(\langle \eta|\eta \rangle -1 \right)  \right] - \langle J||\eta| \rangle \right\}\nonumber \\[1mm]
&&\mbox{\hspace{14mm}} \times \ \exp\left[  - \int_{X_-}^{X_+}  dx\ \! \frac{M}{\beta \hbar^2}\frac{\Delta(x)}{8 \eta^2}\right]\nonumber \\[1mm]
&&\mbox{\hspace{10mm}} -\  \cos(\pi D/2) \int_{\eta(X_-)=-\eta_-}^{\eta(X_+)=\eta_+}  \mathcal{D}\eta(x) \exp\left\{-\beta\left[\langle \eta| \hat{H}| \eta \rangle -\kappa\left(\langle \eta|\eta \rangle -1 \right)  \right] - \langle J||\eta| \rangle \right\}
\nonumber \\[1mm]
&&\mbox{\hspace{14mm}} \times \ \exp\left[  - \int_{X_-}^{X_+}  dx\ \! \frac{M}{\beta \hbar^2}\frac{\Delta(x)}{8 \eta^2}\right],
\label{method-images-1aa}
\end{eqnarray}
where Dirac's notation was employed. A few comments are in order about the right-hand-side of the above relation. First, it should be noticed the presence of the parity-even terms $\langle J||\eta| \rangle$ in PI's. Second, PI's  differ by their respective  Dirichlet boundary conditions.  
Third, the super-index ``$R$'' was used to stress the restricted nature of the fluctuations in the radial PI measure while the measure without ``$R$'' represents a  usual one-dimensional PI measure, i.e.
\begin{eqnarray}
\mathcal{D}^{R}\eta(x) \ \dot{=} \  \lim_{N \rightarrow \infty} \left(\frac{\beta \hbar^2}{2\pi \varepsilon M}  \right)^{\!\!N/2} \prod_{k=1}^{N-1} \int_0^{\infty} d \eta_k, \;\;\;\; \mathcal{D}\eta(x) \ \dot{=} \  \lim_{N \rightarrow \infty} \left(\frac{\beta \hbar^2}{2\pi \varepsilon M}  \right)^{\!\!N/2} \prod_{k=1}^{N-1} \int_{-\infty}^{\infty} d \eta_k\, .
\end{eqnarray}
Here ``$\dot{=}$'' denotes De Witt's ``equivalence'' symbol~\cite{DeWitt}.
Finally, the correct time-sliced form of the exponential with the centrifugal potential is (cf. e.g., Refs.~\cite{Kleinert,Steiner})
\begin{eqnarray}
&&\exp\left[  - \int_{X_-}^{X_+}  dx\ \! \frac{M}{\beta \hbar^2}\frac{\Delta(x)}{8 \eta^2}\right]_{\!R} \ \dot{=} \ \lim_{N\rightarrow \infty}\prod_{k=1}^N \sqrt{2\pi \frac{\beta \hbar^2}{M} \frac{\eta_k\eta_{k-1}}{\varepsilon } \tilde{\Delta}_k} \ \! \exp\left(-\frac{\beta \hbar^2}{M} \frac{\eta_k\eta_{k-1}}{\varepsilon } \tilde{\Delta}_k \right) I_{\frac{D-2}{2}}\left( \frac{\beta \hbar^2}{M} \frac{\eta_k\eta_{k-1}}{\varepsilon }\tilde{\Delta}_k\right), \nonumber \\[2mm] 
&&\exp\left[  - \int_{X_-}^{X_+}  dx\ \! \frac{M}{\beta \hbar^2}\frac{\Delta(x)}{8 \eta^2}\right] \ \dot{=} \ \lim_{N\rightarrow \infty}\prod_{k=1}^N \psi_{\frac{D-2}{2}}\left(-\frac{\beta \hbar^2}{M} \frac{|\eta_k\eta_{k-1}|}{\varepsilon } \tilde{\Delta}_k\right), 
\label{centrifugal-slicing-1a}
\end{eqnarray}
with 
\begin{eqnarray}
\tilde{\Delta}_k \ \equiv \ \tilde{\Delta}(x_k) \ = \ \begin{cases}
  -(D-1)(D-3) & {\rm for~} x_k \in [x_a , x_b] \\
  1 & {\rm otherwise} ,
\end{cases}
\end{eqnarray}
and  (see, e.g.,~Refs.~\cite{Steiner,Dinge})
\begin{eqnarray}
\psi_p(x) \ = \ \frac{ e^x }{\sin(\pi p)} \sqrt{\frac{\pi x}{2}}  \left[I_{-p}(x) - I_{p}(x)  \right]
\ = \ e^x \sqrt{\frac{2 x}{ \pi}} K_p(x)\, .   
\end{eqnarray}
[$I_p$ and $K_p$ are the modified Bessel functions of the first and the second kind, respectively.] In cases when $x \gg 1$ meaning that $|\eta_k \eta_{k-1}| \gg \varepsilon$ (e.g.,  ``typical situation'' for very fine time slicings) the asymptotic form of $\psi_p(-x) \sim 1 + (1-4p^2)/8x + {\mathcal{O}}(1/x^2)$ holds. With help of the preceding asymptotic behavior one obtains
\begin{eqnarray}
\prod_{k=1}^N \psi_{\frac{D-2}{2}}\left(-\frac{\beta \hbar^2}{M} \frac{|\eta_k\eta_{k-1}|}{\varepsilon } \tilde{\Delta}_k \right) \ &\sim& \  \prod_{k=1}^N \left[1- \frac{M}{\beta \hbar^2} \frac{(D-3)(D-1)}{8|\eta_k\eta_{k-1}|\tilde{\Delta}_k} \ \!\varepsilon + {\mathcal{O}}(\varepsilon^2)\right]\nonumber \\[1mm]
&\sim& \  \prod_{k=1}^N \exp\left[- \frac{M}{\beta \hbar^2} \frac{\Delta_k}{8|\eta_k\eta_{k-1}|} \ \!\varepsilon + {\mathcal{O}}(\varepsilon^2) \right]\nonumber \\[1mm]
&\sim& \  \exp\left[  - \int_{X_-}^{X_+}  dx\ \! \frac{M}{\beta \hbar^2}\frac{\Delta(x)}{8 \eta^2}\right]. 
\label{48.aa}
\end{eqnarray}
Potential singularities of the integral at $\eta = 0$ can be regularized, e.g., by a principal value prescription.
Unfortunately, the formula (\ref{48.aa}) cannot be directly used in our case because  the boundary values $\eta_-$ and $\eta_+$ are arbitrarily close to zero, and hence the assumed asymptotic behavior for $\psi_p$ is not fulfilled. This situation can be rectified by factorizing out the problematic boundary points as
\begin{eqnarray}
&&\prod_{k=1}^N \psi_{\frac{D-2}{2}}\left(-\frac{\beta \hbar^2}{M} \frac{|\eta_k\eta_{k-1}|}{\varepsilon } \tilde{\Delta}_k \right) \nonumber\\[2mm] 
&&\sim \  
\frac{1}{\pi}\left(\frac{\beta \hbar^2}{2M \varepsilon}\right)^{D-1} \left(|\eta_-\eta_{1}||\eta_+ \eta_{_{N-1}}|\right)^{(D-1)/2}\ \!
\prod_{k=2}^{N-1} \exp\left[- \frac{M}{\beta \hbar^2} \frac{\Delta_k}{8|\eta_k\eta_{k-1}|} \ \!\varepsilon + {\mathcal{O}}(\varepsilon^2) \right].
\end{eqnarray}
Here we have utilized the asymptotic form  $\psi_p(-x) \sim (-x/2)^{p + 1/2} \ \!\Gamma(-p)/\sqrt{\pi} + \mathcal{O}(x^{p+3/2})$ valid for $x \ll 1$ and $p<0$ ($p\notin \mathbb{Z}^-$).

The passage from the radial PI (\ref{OffDiagRescaled}) to the ordinary (1-dimensional) PI brings about an important advantage, namely, one can perform the WKB approximation. In particular, one can use the Laplace's formula of the asymptotic calculus~\cite{Bleistein,Erdelyi} 
\begin{eqnarray}
\int_{-\infty}^{\infty} dt \ \! f(t)  \exp\left[-\beta g(t)   \right]  \ = \  \sqrt{\frac{2\pi}{\beta g''(t_0)}} \ \! f(t_0) \ \! \exp\left[-\beta g(t_0)   \right] \ + \ {\mathcal{O}}\left( \frac{\exp\left[-\beta g(t_0)   \right]}{\beta^{3/2}} \right),
\label{laplace}
\end{eqnarray}
with $t_0$ being a solution of $g'(t) = 0$ (provided $g(t)$ has a smooth absolute minimum at the interior point $t=t_0~(\neq \pm \infty)$). In case of need, the full asymptotic expansion can be found, e.g., in Ref.~\cite{Erdelyi}. 
In our case the customary PI substitution 
\begin{eqnarray}
\sqrt{\frac{2\pi}{\beta g''(t_0)}} \, \mapsto \, \lim_{N\rightarrow \infty }\sqrt{\frac{\beta \hbar^2}{2\pi
\varepsilon M}} \ \!  \left[{\det}_{_N}\!\left(-\varepsilon^2 \nabla \overline{\nabla} +
\varepsilon^2 \frac{2M}{\hbar^2}(V(x) - E_0) \right) \right]^{-1/2}\, ,
\label{VI.49a}
\end{eqnarray}
should be employed~\cite{Zinn,Kleinert,Grosche}. Here $E_0$ is the ground state energy and the difference operators (lattice derivatives) are defined as~\cite{Kleinert}
\begin{eqnarray}
 \nabla \eta(x) \ = \ \frac{1}{\varepsilon}\left[\eta(x + \varepsilon)  \ - \
 \eta(x) \right]\, , \;\;\;\;\;   \overline{\nabla} \eta(x) \ = \
\frac{1}{\varepsilon}\left[\eta(x)  \ - \
 \eta(x-\varepsilon) \right]\, ,
\end{eqnarray}
with $\nabla \overline{\nabla} = \overline{\nabla} \nabla $ being the Hermitian operator on the space of ``time-sliced'' functions with vanishing end points, i.e., $\eta(x_{N}) = \eta(x_0) = 0$.
The ensuing determinant ${\det}_{_N}(\cdots)$ in (\ref{VI.49a}) can be
conveniently computed by means of the Gelfand--Yaglom method~\cite{G-Y} (or shifting method~\cite{Zinn}). 
There the determinant ${\det}_{_N}(\cdots) = D(x_{N})/\varepsilon$ where $D(x)$ is finite in the $\varepsilon \rightarrow 0$ limit 
and fulfills the differential equation 
\begin{eqnarray}
\left[\frac{d^2}{d x^2}  \ - \ \frac{2M}{\hbar^2} (V(x)-E_0)\right] D(x) \ = \
0\, ,
\label{VI.51a}
\end{eqnarray}
with the Cauchy conditions $D(X_-) = 0$ and $D'(X_-) = 1$. The solution $D(x)$ can be written as a linear combination of two independent solutions of (\ref{VI.51a}). One is immediate, namely the WKB solution $\eta_0(x)$. A second solution can be constructed with the help of d'Alambert's formula  directly from $\eta_0(x)$  in the 
form (cf. e.g., Ref.~\cite{Zettl}) 
\begin{eqnarray}
\eta_0(x) \int^x_{X_-} \frac{d\xi}{\eta_0^2(\xi)} \, . 
\label{VI.52a}
\end{eqnarray}
It is simple to verify that (\ref{VI.52a}) is again a solution of (\ref{VI.51a}). The Cauchy conditions in the Gelfand--Yaglom method uniquely fix 
the constants in front of the two independent solutions so that the full solution $D(x)$ acquires the form
\begin{eqnarray}
D(x) \ = \  \eta_0(x) \eta_0(X_-) \int^x_{X_-} \frac{d\xi}{\eta_0^2(\xi)} 
\ = \  \eta_0(x) \eta_- \int^x_{X_-} \frac{d\xi}{\eta_0^2(\xi)} \, .
\end{eqnarray}
As a result we obtain in the large $N$ (i.e., small $\varepsilon$) limit
\begin{eqnarray}
{\det}_{_N}\!\left[-\varepsilon^2 \nabla \overline{\nabla} +
\varepsilon^2 \frac{2M}{\hbar^2}(V(x)-E_0) \right] \ \stackrel{\varepsilon \rightarrow 0}{\longrightarrow} \  \frac{\eta_+ \eta_-}{\varepsilon}  \int^{X_+}_{X_-} \frac{dx}{\eta_0^2(x)}\, .
\label{VI.55aa}
\end{eqnarray}
The great advantage of the Gelfand--Yaglom method is that it does not require any detailed knowledge of the spectrum of the operator 
whose determinant is being computed. It also shows that the determinant in (\ref{VI.49a}) diverges linearly with $N$ in the large $N$ limit. 

We now substitute for $f(t_0)$ in (\ref{laplace}) the functional expression
\begin{eqnarray}
 \exp\left[ - \langle J||\eta_0| \rangle  - \int_{X_-}^{X_+}  dx\ \! \frac{M}{\beta \hbar^2}\frac{\Delta(x)}{8 \eta^2_0}\right] \, ,
\end{eqnarray}
where, $\eta_0(x)$ comes from the path that minimizes the functional $\langle \eta| \hat{H}| \eta \rangle -\kappa\left(\langle \eta|\eta \rangle -1 \right)$. According to the  Rayleigh--Ritz variation principle (see, e.g., Refs.~\cite{Messiah,Griffiths}), such a function $\eta(x)$ is the ground-state wavefunction of the Hamiltonian $\hat{H}$, i.e., $\eta_0(x) = \psi_{gs}(x)$ with $\kappa = E_0$.    
With a hindsight, we have used the latter value of $\kappa$ already in Eqs.~(\ref{VI.49a}), (\ref{VI.51a}) and (\ref{VI.55aa}). 
Notice that the stationary point in $\kappa$ is real but the integration contour in $\kappa$ is parallel to the imaginary axis.
Both reality and positivity  of $\eta(x)$ pose no restriction in the Rayleigh--Ritz principle, because the ground state can always be chosen real and positive~\cite{MurPetit}. 
Putting it all together, we get the leading large-$\beta$ behavior in the form
\begin{equation} \label{SpectralAsymp2a}
\rho(x_a,x_b,\beta) \ = \ 
e^{-\beta E_{gs}}  \ \! \psi_{gs}(x_a) \psi_{gs}(x_b)\, ,
\end{equation}
as it should be (cf. Eq.~(\ref{SpectralAsymp2})).

We conclude the discussion of low-temperature expansion by noting that the Rayleigh--Ritz variation principle
states that {\em all} eigenvalues and (normalized) eigenvectors of $\hat{H}$
come from stationary solutions of  $\langle \eta| \hat{H}| \eta \rangle
-\kappa\left(\langle \eta|\eta \rangle -1 \right)$, and
conversely~\cite{Messiah}. In the spirit of the WKB approximation one should sum over 
all path integrals evaluated about {\em all} stationary solutions. It is, however, only the ground state configuration
$\{\psi_{gs}(x),E_0\}$ that acquires the global minimum and which gives the larges contribution to the WKB approximation. 
This fact was implicitly used in our preceding reasonings. Should we have included also other 
stationary solutions we would recover higher order terms in the spectral expansion of the Bloch density matrix (\ref{SpectralAsymp2}).
\section{Conclusion and outlook}
\label{sec:Conclusion}

In this paper, we have derived the local-time PI representation of the Bloch density matrix. 
We have shown that the result obtained, apart from being of interest in pure mathematics (stochastic theory, Sturm--Liouville theory, etc), 
can serve as a useful alternative to the traditional Feynman--Kac PI representation of Green functions of
Fokker--Planck equations. Furthermore, by analytically continuing the result back to the real time via the inverse Wick rotation, $\beta \rightarrow i t/\hbar$, one obtains the local-time PI representation of quantum-mechanical transition amplitudes, i.e., matrix elements of the evolution operator $e^{-i t \hat{H}/\hbar}$. From the physicists' point of view, perhaps the most important application of local-time PI's lies in statistical physics, and namely in the low and high temperature  treatments of the Bloch density matrix. This is because 
in conventional PI's  only a very tiny subset of paths gives a relevant contribution in these asymptotic regimes. 
In particular, the high-temperature regime of the Boltzmann density function $\rho(x,x,\beta)$ is dominated by paths that spend a 
sizable amount of time in the vicinity of the point $x$. Similarly, the low-temperature regime is controlled by paths with 
a large local time near the global minimum of the potential. Here we have exemplified the conceptual convenience of the local-time formulation by providing a generic analysis of the low-temperature behavior of the Bloch density matrix. Our formulation proved to be particularly instrumental in obtaining the correct asymptotic behavior (known from the spectral theory) which is otherwise notoriously difficult to obtain within the Feynman--Kac PI framework~\cite{FeynmanHibbs,Paulin}.  As a byproduct we have uncovered an interesting connection between a low-temperature PI expansion, the Gelfand--Yaglom formula and the Rayleigh--Ritz variational principle. 

In order to further reinforce our analysis, we formulated 
a local-time analog of the Feynman--Matthews--Salam formula which is (similarly as its QFT counterpart) expedient in number of statistical-physics contexts.  The prescription obtained was substantiated by an explicit calculation of a one-point distribution function of the local time.  

It appears worthwhile to stress that our local-time representation (with its build-in replica field trick) is in its present form applicable only to one-dimensional quantum mechanical systems. With a hindsight we reflected this fact already in our choice of the incipient PI (\ref{PIx}) where we assumed $\tau \in \mathbb{R}$ and $x \in \mathbb{R}$. Though one may easily proceed up to Eq.~(\ref{ReplRepre}) without any restriction on the value of $d$ (in fact, Eq.~(\ref{ReplRepre})  is valid for any $x \in \mathbb{R}^d$ with $d \geq 1$), a further progress in this direction is hindered by the fact that the replica fields depend on a $d$-dimensional argument $x$, 
and thus the PI in (\ref{ReplRepre}) can no longer be regarded as a quantum mechanical  PI (i.e., PI over fluctuating paths). In effect, we cannot use existing mathematical techniques of the PI calculus  (e.g., transformation of PI's to polar coordinates), that we have employed to get the radial PI~(\ref{DiagLapl}). 
The issue of the extension of our local-time PI representation to higher-dimensional configuration space is currently under active
investigation.
%

\section*{Acknowledgements}

The authors acknowledge A.~Alastuey, H.~Kleinert, Z.~Haba and D.S.~Grebenkov for fruitful discussions. The work has been supported by 
the GA\v{C}R Grant No. GA14-07983S. V.Z. was also partially supported by the CTU in Prague Grant No. SGS13/217/OHK4/3T/14 and by 
the DFG Grant: KL 256/54-1.

\appendix
\section{Off-diagonal matrix elements}
\label{sec:App}

In this Appendix we show that the representation (\ref{LaplOffDiag}) reduces to the well established result (\ref{SturmGreen}) of the Sturm--Liouville theory. Since $\widetilde{\rho}(x_a,x_b,E)$ is symmetric in $x_a$ and $x_b$, we will assume, without loss of generality, that $x_a < x_b$.

The path integral in (\ref{LaplOffDiag}) can be expressed via Eq.~(\ref{RadialHOSol}) as
\begin{equation}
\int_0^\infty \!\! d\eta_a d\eta_b \ \! \ \! (\eta_+ X_+ | \eta_b x_b )_{D} \eta_b (\eta_b x_b | \eta_a x_a )_{2} \eta_a (\eta_a x_a | \eta_- X_- )_{D}\, .
\end{equation}
The limits in Eq.~(\ref{LaplOffDiag}) can be carried with the help of the asymptotic formulas $I_{D/2-1}(z)~\approx~(z/2)^{D/2-1} / \Gamma(D/2)$, and $\Gamma(z)~\approx~1/z$, valid for $z \rightarrow 0$. We obtain
\begin{eqnarray} \label{etaLimMinus}
\frac{1}{D} \eta_-^{\frac{1-D}{2}} (\eta_a x_a | \eta_- X_- )_{D}
&& \xrightarrow{\eta_- \rightarrow 0} \ 
\left( \frac{\hbar^2 \eta_a}{2 M F_1(x_a)} \right)^{\!\! D/2}
\frac{\exp \left( -\frac{\hbar^2}{2M} \frac{F'_1(x_a)}{F_1(x_a)} \eta_a^2 \right)}{\frac{D}{2} \Gamma(\frac{D}{2}) \sqrt{\eta_a}}
\nonumber\\[2mm]
&& \xrightarrow{D \rightarrow 0} \ 
\frac{1}{\sqrt{\eta_a}}
\exp \left( -\frac{\hbar^2}{2M} \frac{F'_1(x_a)}{F_1(x_a)} \eta_a^2 \right)\! ,
\end{eqnarray}
where $F_1(x)$ satisfies Eq.~(\ref{DiffEqFG}) with initial conditions $F_1(X_-) = 0$ and $F'_1(X_-) = 1$, and
similarly, we find
\begin{equation} \label{etaLimPlus}
\frac{1}{D} \eta_+^{\frac{1-D}{2}} (\eta_+ X_+ | \eta_b x_b )_{D}
\ \xrightarrow{\eta_+ \rightarrow 0, D \rightarrow 0}\
\frac{1}{\sqrt{\eta_b}}
\exp \left( \frac{\hbar^2}{2M} \frac{G'_3(x_b)}{G_3(x_b)} \eta_b^2 \right)\, ,
\end{equation}
where $G_3(x)$ satisfies Eq. (\ref{DiffEqFG}) with initial conditions $G_3(X_+) = 0$ and $G'_3(X_+) = -1$.
Formula (\ref{LaplOffDiag}) then reduces to
\begin{eqnarray}
\widetilde{\rho}(x_a,x_b,E) &=&
\frac{2 \hbar^2}{M} \int_0^\infty  d\eta_a d\eta_b \ \!
\frac{\eta_a \eta_b}{G_2(x_a)} \ \! I_{0}\!\left(\frac{\hbar^2}{M} \frac{\eta_a \eta_b}{G_2(x_a)}\right)\nonumber\\[2mm]
&&\times \ \!
 \exp \left[ - \frac{\hbar^2}{2M} \left(
\frac{W(G_2,F_1) \eta_a^2}{F_1(x_a) G_2(x_a)} +
\frac{W(G_3,F_2) \eta_b^2}{F_2(x_b) G_3(x_b)} \right)
\right]\, ,
\end{eqnarray}
where $F_2(x)$ and $G_2(x)$ satisfy Eq. (\ref{DiffEqFG}) with initial conditions $F_2(x_a) = 0$ and $F'_2(x_a) = 1$, and $G_2(x_b) = 0$ and $G'_2(x_b) = -1$, respectively. The Wronskian $W(F,G) \equiv F(x) G'(x) - F'(x) G(x)$ is independent of $x$, as discussed in Section \ref{sec:Radial}, and antisymmetric, i.e., $W(F,G) = - W(G,F)$.

Wronskians $W(G_2,F_1)$ and $W(G_3,F_2)$ assume a particularly simple form when evaluated  at points $x_b$ and $x_a$, respectively, due to the initial conditions satisfied by $G_2$ and $F_2$. We find $W(G_2,F_1) = F_1(x_b)$ and $W(G_3,F_2) = G_3(x_a)$. Rescaling $\eta_a \rightarrow \sqrt{G_2(x_a) M / \hbar^2}\eta_a$, $\eta_b \rightarrow \sqrt{F_2(x_b) M / \hbar^2}\eta_b$, and using the relation $F_2(x_b) = G_2(x_a)$ we obtain
\begin{equation}
\widetilde{\rho}(x_a,x_b,E) \ = \
\frac{2 M}{\hbar^2} G_2(x_a) \int_0^\infty \! d\eta_a d\eta_b \ \!
\eta_a \eta_b \ \! I_{0} (\eta_a \eta_b)
\exp \left(
- \frac{F_1(x_b)}{2 F_1(x_a)} \eta_a^2
- \frac{G_3(x_a)}{2 G_3(x_b)} \eta_b^2 \right) \! .
\end{equation}
The integrations are readily performed using the formula \cite{Gradshteyn}
\begin{equation}
\int_0^\infty dz \ \! I_0(b z) \exp\left(-\frac{a}{2} z^2\right) \ = \
\frac{1}{a} \exp\left(\frac{b^2}{2a}\right)\, ,
\end{equation}
yielding
\begin{equation}
\widetilde{\rho}(x_a,x_b,E) \ = \
\frac{2 M}{\hbar^2}
\frac{F_1(x_a) G_3(x_b) G_2(x_a)}{F_1(x_b) G_3(x_a) - F_1(x_a) G_3(x_b)}\, .
\end{equation}
To prove equality with (\ref{SturmGreen}), we only have to show that
\begin{equation}
F_1(x_a) G_3(x_b) - F_1(x_b) G_3(x_a) \ = \ G_2(x_a) W(F_1,G_3)\,  .
\end{equation}
This is done by realizing that $G_2(x)$, being a solution of the second-order linear differential equation (\ref{DiffEqFG}), can be uniquely composed as a linear combination of two other solutions $F_1(x)$ and $G_3(x)$,
\begin{equation}
G_2(x) \ = \ \frac{F_1(x) G_3(x_b) - F_1(x_b) G_3(x)}{W(F_1,G_3)}\, .
\end{equation}
Indeed, thus defined $G_2$ satisfies the initial conditions $G_2(x_b) = 0$ and $G'_2(x_b) = -1$.

We conclude that
\begin{equation}
\widetilde{\rho}(x_a,x_b,E) \ = \
- \frac{2 M}{\hbar^2}
\frac{F_1(x_a) G_3(x_b)}{W(F_1,G_3)}\, ,
\end{equation}
which coincides with the Sturm--Liouville result (\ref{SturmGreen}).


\end{document}